\documentclass[conference]{IEEEtran}
\IEEEoverridecommandlockouts

\IEEEpubid{\parbox{\columnwidth}{\scriptsize \copyright{} 2026 IEEE. Personal
use of this material is permitted. Permission from IEEE must be obtained for all
other uses, in any current or future media, including reprinting/republishing
this material for advertising or promotional purposes, creating new collective
works, for resale or redistribution to servers or lists, or reuse of any
copyrighted component of this work in other works.\hfill}\hspace{\columnsep}\makebox[\columnwidth]{}}

\usepackage{booktabs}
\usepackage{fancyvrb}
\usepackage{cite}
\usepackage{graphicx}
\usepackage{url}

\newcommand{\Description}[1]{}

\title{The Polyglot's Dilemma: Conformance Testing a Dozen Specs in as Many Languages}

\author{
\IEEEauthorblockN{A. Jesse Jiryu Davis}
\IEEEauthorblockA{MongoDB Research\\
New York, New York, USA\\
jesse@mongodb.com}
\and
\IEEEauthorblockN{Jeremy Mikola}
\IEEEauthorblockA{Independent\\
Hoboken, New Jersey, USA\\
jmikola@gmail.com}
\and
\IEEEauthorblockN{Jeff Yemin}
\IEEEauthorblockA{MongoDB, Inc.\\
New York, New York, USA\\
jeff.yemin@mongodb.com}
}

\begin{document}

\maketitle

\begin{abstract}
MongoDB maintains client libraries in a dozen programming languages, used by tens of thousands of organizations and millions of developers. Most are implemented natively rather than as wrappers around a shared core. Ensuring consistent behavior across these libraries, comprising millions of lines of code, is hard but essential. Over eleven years, we developed a specification-based testing approach: tests are written once in YAML and executed by language-specific interpreters for each library. We describe the evolution from many ad-hoc formats to a Unified Test Format, which allowed us to delete over 22,000 lines of test code. The rate of nonconformance bugs fell up to 86\% in drivers that adopted YAML tests (though results varied). We report lessons learned about declarative test design, test architecture, schema evolution, and the limits of unification.
\end{abstract}

\begin{IEEEkeywords}
NoSQL, MongoDB, protocol testing, conformance testing
\end{IEEEkeywords}

\section{Introduction}
\IEEEpubidadjcol

Database client libraries are often built as thin wrappers around a shared core written in C. This architecture minimizes duplication, but shared-core libraries impose costs on users. Compiling and installing a C dependency is unfamiliar for many programmers, the library cannot participate in language-native async event loops, and its API rarely feels idiomatic.

At MongoDB, we chose a different path. We maintain a dozen client libraries (the number has changed over time), which we call ``drivers.'' We wrote the entire driver logic and API nine times in C, C\#, Go, Java, Node.js, Python, Ruby, Rust, and Swift. In addition, we wrote idiomatic layers in C++ and PHP that wrap the C Driver, and in Scala and Kotlin wrapping the Java Driver. This approach lets each driver integrate naturally with its ecosystem---it is easy to install, obeys native logging conventions, cooperates with async event loops, and presents an API that feels familiar to developers in that language. The cost is that we maintain the same client-server protocol, connection pooling, server discovery, and other behaviors in at least nine code bases.

Prose specifications help, but natural language is ambiguous. Differential testing~\cite{mckeeman_differential_1998} can detect how implementations diverge, but it cannot say which implementation is correct. It also cannot detect when all implementers make the same mistake---so-called common-mode failures---because the spec is unclear or because some errors are easy to make~\cite{brilliant_consistent_1987}.

Our solution is a \emph{domain-specific test language}, or DSTL: we write machine-executable tests in YAML that encode the spec author's intent. Each driver implements a test runner that interprets these YAML files, runs the specified operations against a MongoDB deployment, and asserts that the results match expectations. When specs change, we update the YAML tests in a central repository, and the changes propagate automatically to the individual drivers.

Starting in 2015, we developed and refined this approach. Our early efforts produced multiple test formats, each tailored to a particular specification: one DSTL for create, read, update, delete (CRUD) operations, another for transactions, another for connection pooling, and so on. By 2020, this proliferation had become a maintenance burden. We consolidated these formats into a Unified Test Format (UTF), validated by a JSON Schema. By migrating to UTF we deleted over 22,000 lines of redundant test-runner code, while gaining schema validation and a more expressive test language.

UTF introduced several mechanisms that we will explain in detail below. An \emph{entity map} is a registry of objects---clients, databases, collections (i.e. database tables), cursors---created during a test, allowing operations to reference them by name later in the test. \emph{Command monitoring} captures every message the driver sends to the server and every reply it receives, enabling tests to check wire-level behavior. \emph{Fail points} are server-side hooks that simulate errors, dropped connections, and slow responses, providing controlled conditions for testing error handling.

This paper reports on eleven years of industrial experience applying this methodology across a dozen driver implementations, comprising millions of lines of production code. Our drivers are used by tens of thousands of customers and millions of open source users. Even subtle inconsistencies in their behaviors are costly. While our techniques are not novel compared to recent research prototypes, we offer something complementary: empirical evidence that declarative, cross-language conformance testing scales to production systems over extended timeframes. We report what worked, what did not, and where we reached the limits of unification.

\begin{figure}[ht]
    \includegraphics[width=0.9\linewidth]{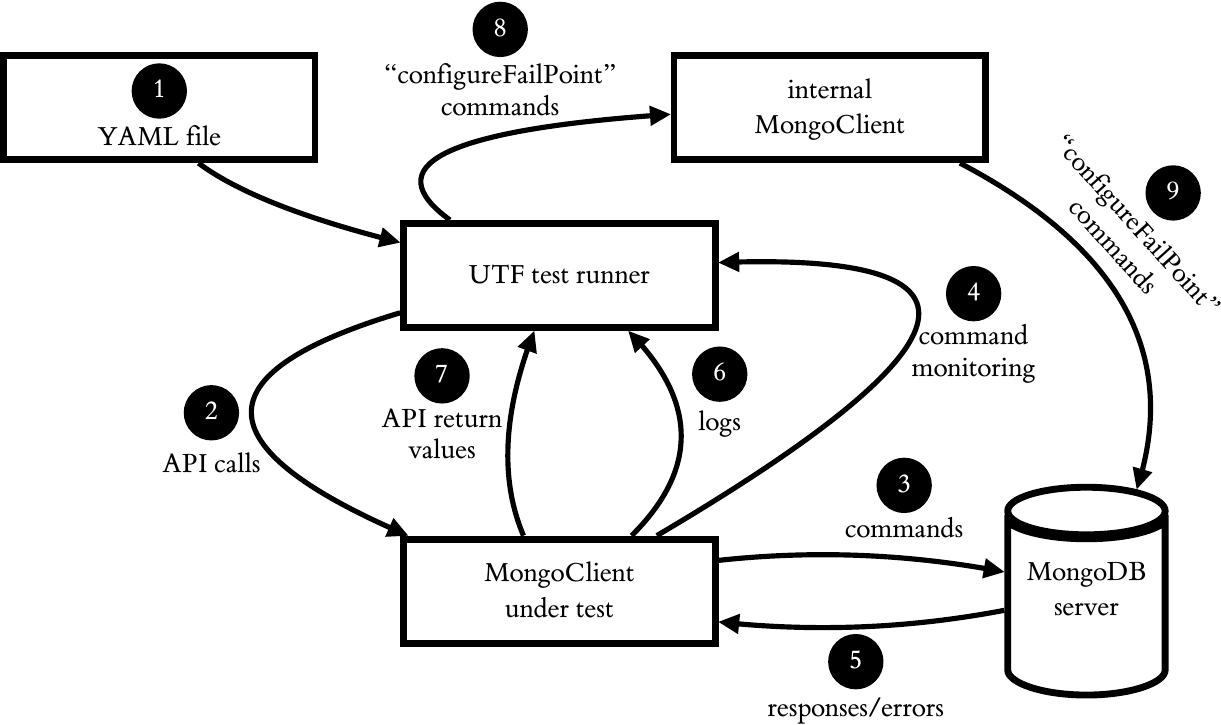}
    \caption{MongoDB driver specification testing architecture.}
    \label{fig:architecture}
    \Description{A flowchart showing control starting at a YAML file, flowing into a UTF test runner, and then flowing among other components.}
\end{figure}

\section{Test Architecture and Components}

The components for testing a MongoDB driver in any language with UTF are shown in Figure~\ref{fig:architecture}. The UTF test runner reads a YAML file (1), and creates two client objects (MongoClients): a MongoClient under test, and an internal MongoClient for out-of-band communications with the MongoDB server. The test runner executes the YAML file's list of operations. For each operation it calls a method on the MongoClient under test (2), which sends commands to the server (3), and also reports them back to the test runner (4) via the command monitoring API. The server responds to the client (5), which may log to a file-like stream created by the test runner (6), and finally the original API call returns a value or error (7). The YAML file's operations may include fail points that the test runner should enable or disable to control the server's behavior; these are interleaved with other test operations. The test runner sends \texttt{configureFailPoint} commands to the server via its internal MongoClient to avoid interference with the MongoClient under test. The test runner runs one command at a time, synchronously, on either MongoClient, to ensure tests are deterministic.

We originally designed the command monitoring API for application performance monitoring (APM), but we realized we could use it to test drivers' network messages. Fail points are also vital: we can command the server to misbehave in various ways, e.g., return an error the next \textit{N} times a certain command is called, or block a command for a certain period, or close the connection. Compared to stress testing or ``chaos'' testing, our approach is fast, deterministic, and simple, while still exercising drivers' error paths. We use APM and fail points to test behavior that is invisible from the API surface, as in the following example.

Figure~\ref{fig:utf-example} shows a complete test file in UTF. The test creates a client, database, and collection as named \textit{entities}, referenced by later operations using YAML anchors. The test populates a collection with initial data and tells the server to close the connection on the next \texttt{update} command. The runner calls the client's \texttt{updateOne} API. The test runner uses APM to check that the driver sends exactly two \texttt{update} commands. It uses \texttt{waitForEvent} to asynchronously check that the driver eventually marks the server's type as \texttt{Unknown} after the network error, meaning the driver will re-check which server is the primary. Without APM and fail points these checks would be inconvenient or unreliable. Finally, \texttt{expectResult} asserts that the API call succeeded, and \texttt{outcome} confirms the document was updated once.


\begin{SaveVerbatim}{utfExample}
schemaVersion: "1.4"
runOnRequirements:
  - {minServerVersion: "4.4", topologies: [replicaset]}
createEntities:
  - client:
      id: &client client0
      observeEvents:
        - commandStartedEvent
        - poolClearedEvent
        - serverDescriptionChangedEvent
  - database:
      id: &db db0
      client: *client
      databaseName: &dbName mydb
  - collection:
      id: &coll coll0
      database: *db
      collectionName: &collName mycoll
initialData:
  - collectionName: *collName
    databaseName: *dbName
    documents: [{_id: 1, x: 1}]
tests:
  - description: "updateOne retries after network error,
                  exactly-once execution"
    operations:
      - name: failPoint
        object: testRunner
        arguments:
          client: *client
          failPoint:
            configureFailPoint: failCommand
            mode: {times: 1}
            data:
              failCommands: [update]
              closeConnection: true
      - name: updateOne
        object: *coll
        arguments:
          filter: {_id: 1}
          update: {$inc: {x: 1}}
        expectResult:
          matchedCount: 1
          modifiedCount: 1
          upsertedCount: 0
      - name: waitForEvent
        object: testRunner
        arguments:
          client: *client
          event:
            serverDescriptionChangedEvent:
              newDescription: {type: Unknown}
          count: 1
    expectEvents:
      - client: *client
        eventType: command
        events:
          - commandStartedEvent: {commandName: update}
          - commandStartedEvent: {commandName: update}
      - client: *client
        eventType: cmap
        events: [{poolClearedEvent: {}}]
    outcome:
      - collectionName: *collName
        databaseName: *dbName
        documents: [{_id: 1, x: 2}] # x incremented once
\end{SaveVerbatim}
\begin{figure}[ht]
\setlength{\fboxsep}{4pt}
\begin{minipage}{\columnwidth}
\fbox{\BUseVerbatim[fontsize=\scriptsize]{utfExample}}
\end{minipage}
\caption{A UTF test file: entity definitions, initial data, and test operations.}
\label{fig:utf-example}
\end{figure}

\section{Evolution of the Unified Test Format}

\subsection{Choosing YAML for Writing Specification Tests}

We first introduced cross-driver test files with our Server Discovery and Monitoring (SDAM) specification in 2014. These tests checked drivers' behavior when connecting to a MongoDB cluster and responding to topology changes (e.g. a change in the number of servers in a \textit{replica set}, MongoDB's consensus group~\cite{schultz_design_2022}). A test runner fed simulated server messages to a driver's internal state machine and asserted on its state transitions.

We considered the Cucumber testing framework~\cite{wynne_cucumber_2017}, with its English-like Gherkin syntax. But Gherkin is intended to express business logic to non-programmers. Describing binary formats and TCP/IP connections was inconvenient and silly-looking; our engineers rebelled at the proposal. We settled on YAML for its programmer-friendly syntax, support for comments, and the availability of parsing libraries in most target languages. YAML allows basic factoring with \textit{anchors} and \textit{aliases}. Test files could be converted to JSON for languages that lack a YAML parser. The payload of most MongoDB wire protocol messages is BSON, a JSON-like binary format. Since YAML is a superset of JSON, payloads could be naturally embedded in YAML files. In 2017, we developed a MongoDB Extended JSON format that is valid JSON and converts losslessly to and from BSON; e.g., a BSON 64-bit integer is expressed as \texttt{\{"\$numberLong": "42"\}}.

These first YAML tests and all the drivers' test runners were specific to the SDAM spec. We began to proliferate YAML test formats for other specs. By 2015 we had various DSTLs and test runners for driver functionality ranging from connection string parsing, to BSON serialization, to server selection. For the CRUD spec, which defined APIs for read and write operations, we introduced a DSTL called CRUD v1, which would be the foundation of UTF.

\subsection{From CRUD v1 to the Unified Test Format}

The initial CRUD v1 format allowed a driver to execute one operation and assert either a result or an error, and it could check the final state of the database. In 2015 we specified a \textit{command monitoring} API for drivers, allowing applications to observe every message between the driver and the server. This permitted us to extend CRUD v1 with an \texttt{expectEvents} field containing a list of messages and other events to match. In 2017, MongoDB introduced idempotent retry of failed operations. To test this, we began using the server's \texttt{configureFailPoint} command systematically in driver testing---our YAML format could instruct the test runner to set a fail point before running a command. Our multiple test formats had become very powerful, but poorly factored and unmaintainable.

UTF debuted in 2020 with a reenvisioned DSTL, backed by a JSON schema and a thoroughly specified test runner. It superseded all previous formats. At first we manually ported a sample of old tests to prove UTF covered all our requirements. Once UTF was formalized, we wrote scripts for each legacy test format to mechanically convert it to UTF.

There are 28 revisions to the UTF schema to date. Old files validate with the latest schema, and the newest test runners can interpret the oldest files. Each UTF file includes a top-level \texttt{schemaVersion} field, which specifies the JSON schema version to which the file conforms (when mechanically converted from YAML to JSON). This serves two purposes. First, a test runner can determine at a glance if it is capable of interpreting the file. Second, the CI tests for the specifications repository can validate each test file against its schema.

UTF files include a \texttt{runOnRequirements} array containing one or more sets of conjunctive criteria. If at least one set of criteria is satisfied, the test runs:

\begin{verbatim}
# Txns were first implemented for replica
# sets only, then sharded clusters.
runOnRequirements:
  - { minServerVersion: '4.0',
      topologies: [replicaset] }
  - { minServerVersion: '4.2',
      topologies: [sharded] }
\end{verbatim}

The UTF runner maintains a general-purpose object registry: the entity map. Entities are defined in the \texttt{createEntities} section and referenced by name. The \texttt{saveResultAsEntity} syntax saves a command result to the entity map for later use. This permits us to create and reuse any number of collections, documents, etc. during the test.

UTF specifies rules for matching logic in great detail, including: numeric comparisons; selectively ignoring optional fields in BSON documents; ignoring key order when comparing documents; and special operators for complex assertions, e.g., asserting whether a field exists, asserting its type, matching against a dynamic value stored in the entity map.

UTF is still evolving. Some syntax has been deprecated, but none has been removed. In addition to command monitoring, drivers support event-based monitoring of server discovery (SDAM) and connection pooling. A forward-thinking design in \texttt{expectEvents} required minimal changes when incorporating these new event types. When we added concurrent SDAM testing to UTF, we defined a new ``thread'' entity type and associated operations (\texttt{runOnThread}, \texttt{waitOnThread}), reusing the existing \texttt{operations} syntax. Testing for standardized logging warranted a new \texttt{expectLogMessages} field, closely modeled after \texttt{expectEvents}. We used a similar approach for \texttt{expectTracingMessages} to test OpenTelemetry integration.

\subsection{New Tests and Specs}

The requirements for MongoDB drivers are defined in dozens of specifications. When we create or update a spec and its UTF tests, we determine whether we need new UTF syntax or test runner behavior. If so, we change the spec, UTF, and the test runners simultaneously. We usually prototype spec changes in at least two drivers before general publication.

Historically, MongoDB engineers manually synced YAML tests from the specs repository into drivers' repositories. Beginning in 2024, several drivers transitioned to including the specs repository as a Git submodule. We configured GitHub's Dependabot to automatically create pull requests to update the submodule weekly. If CI reports that the driver's test suite passes with the test changes, the pull request is automatically merged. Since test runners skip files with an unsupported \texttt{schemaVersion}, there is generally no issue with updating test files before a driver can implement the latest specs and test format changes.

\subsection{Testing the UTF Test Runner and Its Tests}

The UTF test runner has its own set of tests, which are divided into three categories: ``invalid'' test files are expected to fail schema validation, ``valid-fail'' test files should pass schema validation but fail at runtime (e.g., an assertion failure), and ``valid-pass'' tests should pass without error. Collectively, these tests can assert correctness of the JSON schema. The set of ``valid-fail'' and ``valid-pass'' tests can additionally be used to verify correctness of a driver's test runner. Automated checks on GitHub pull requests validate YAML files against their declared JSON schema version, catching mistakes that were common prior to UTF. The individual drivers' test runners also cross-validate the tests: a failure indicates there is a bug \textit{somewhere}, whether in the test, the driver, or the runner.

\section{Results}

\begin{figure}[ht]
    \centering
    \includegraphics[width=\linewidth]{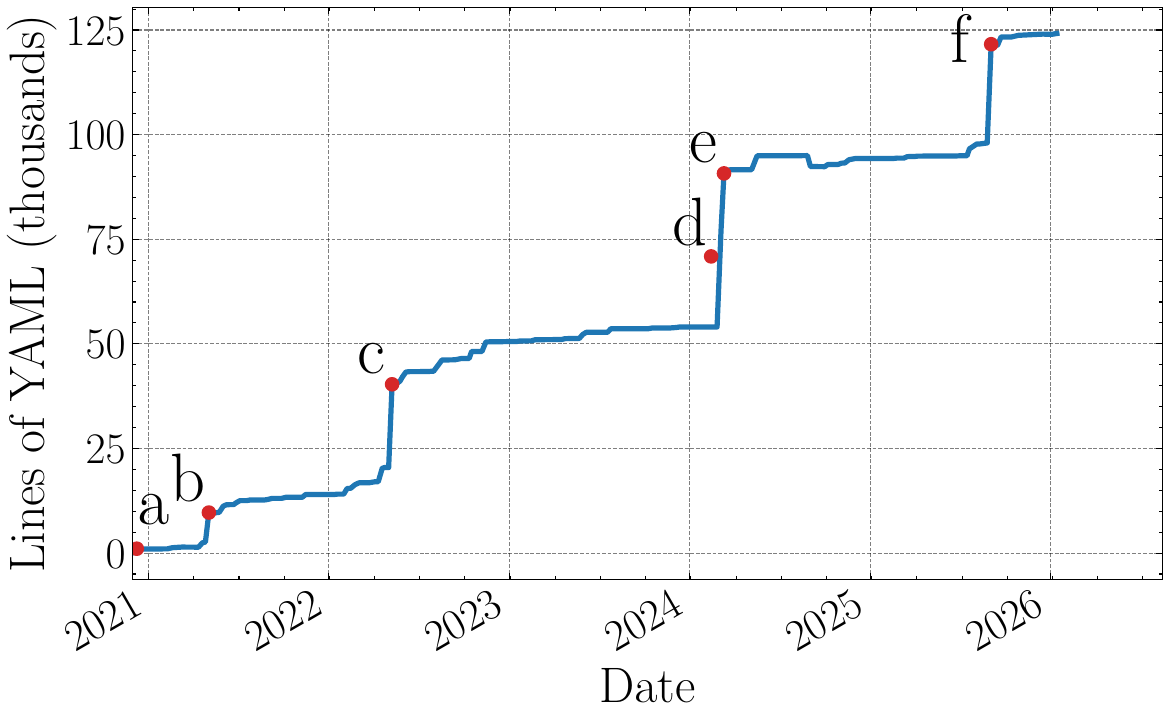}
    \caption{UTF corpus growth.}
    \label{fig:count_utf_lines}
\end{figure}

\begin{figure}[ht]
    \centering
    \includegraphics[width=\linewidth]{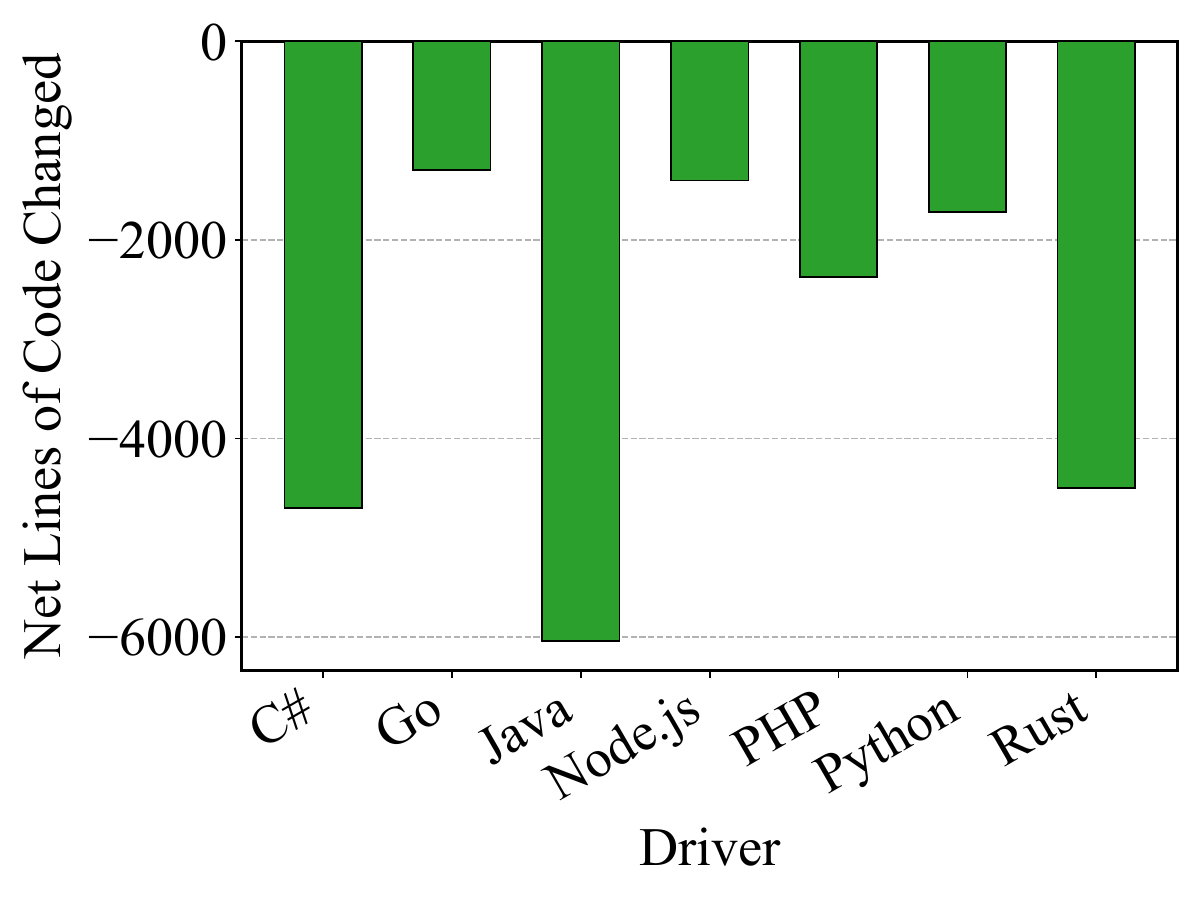}
    \caption{Lines deleted during UTF migration.}
    \label{fig:utf_driver_code_changes}
\end{figure}

Since 2020 our corpus of UTF tests has grown from a prototype with 7 files and 1046 lines of non-comment non-whitespace YAML, to 606 files with 124,168 lines. Figure~\ref{fig:count_utf_lines} shows the major events in UTF's growth: we created UTF (a), ported some CRUD tests (b), added Client-Side Operations Timeout tests (c), ported transactions tests (d), ported remaining CRUD and retryable writes/reads tests (e), and ported Client-Side Field-Level Encryption (f). Most drivers have replaced most of their runners for the various test formats in favor of one UTF runner, and they deleted test code on net. The Java driver had the most dramatic savings of 6038 LOC (Figure~\ref{fig:utf_driver_code_changes}).

\subsection{Bug Prevention}

\begin{figure}[ht]
    \centering
    \includegraphics[width=\linewidth]{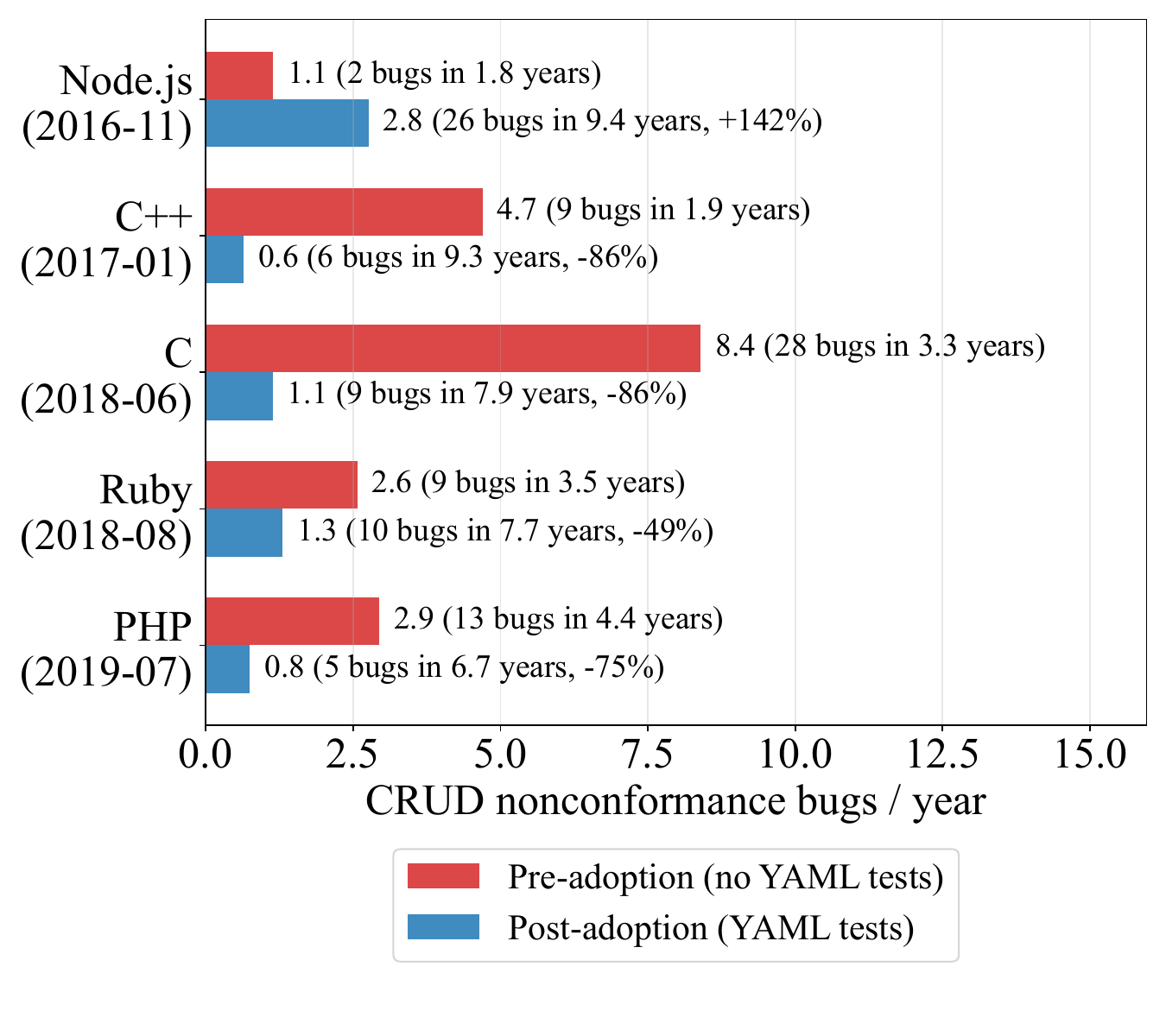}
    \caption{CRUD nonconformance bug rates before and after YAML test adoption (5 late-adopting drivers). Pre-adoption window: spec published, no YAML tests. Post-adoption window: YAML tests adopted.}
    \label{fig:crud_late5}
\end{figure}

Once a spec is published, do the drivers that implement its YAML tests have fewer nonconformance bugs? We analyzed the CRUD spec (one of our largest and oldest specs) and the five drivers with long gaps between the spec's publication in February 2015 and their adoption of its YAML tests. (The other drivers adopted the YAML tests almost immediately, leaving no time to measure their conformance \emph{without} YAML tests.) We used Claude Sonnet to classify 6836 resolved tickets in the five late-adopting drivers as either CRUD nonconformance bugs or other bugs. After review by Claude Opus and a human, 125 CRUD nonconformance bugs remained. The classification prompts, validation data, and analysis pipeline are in our public artifact repository.\footnote{\url{https://github.com/mongodb-labs/issre-2026-polyglots-dilemma-artifacts}} Figure~\ref{fig:crud_late5} shows nonconformance bug rates (bugs/year) before and after YAML test adoption. (The tests were ported from the CRUD-specific format to UTF during our analysis period, but UTF had no effect on the tests' efficacy, only on their maintainability.) Four drivers show reductions of 49--86\%. We found many bugs in the C Driver early, when other drivers that wrap it implemented the spec and the YAML tests.

Even after we adopted YAML tests in these five drivers we found nonconformance bugs, because of coverage gaps in the YAML tests, or bugs in drivers' test runners. This was particularly true of the Node.js driver, where the bug rate was 142\% higher after adoption. Although it passed the YAML tests, its bulk write implementation had many inconsistencies with the spec.

\textbf{Threats to validity:} This is a retrospective study with no control arm. As the drivers matured we would expect bug rates to decline regardless of YAML adoption. There were just a handful of nonconformance bugs per driver, so a few misclassifications would have a large effect. We mitigated this with multiple review passes (see the artifact repository).

\subsection{Lessons Learned}

\textbf{We cannot unify everything:} UTF is not suitable for every spec. We still have specialized YAML formats and runners for testing BSON serialization, the SDAM state machine, connection string parsing, and other features. Some tests are simply described in English. BSON and connection string tests assert on in-process data structures rather than on API calls or wire messages, so the entity map / operation / \texttt{expectEvents} structure does not fit. SDAM tests inject simulated server state directly into a driver's internal state machine, bypassing the public API layer entirely. If our YAML format were capable of describing any possible test, it would have become a Turing-complete programming language with interpreters in a dozen other languages, and its cost and risks would outweigh the benefits. We stopped folding formats into UTF when we reached the point of diminishing returns.

\textbf{Single schema version:} UTF's monotonically increasing schema version led to situations where a driver was blocked from implementing the tests for a new spec, because it had fallen several schema versions behind. If UTF files instead had a list of required test runner features, driver engineers would have had more flexibility about the order in which they implemented specs.

\textbf{Judicious enforcement:} Despite spec authors' efforts, our drivers still have some implementation-specific behavior. For example, a field on a result/response object might be optional or absent in some drivers. We added the \texttt{\$\$unsetOrMatches} operator to UTF so that test authors can permit drivers \textbf{not} to return a value in some cases, but require a \textbf{specific} return value if there is one.

\textbf{Manual test authoring:} Our YAML tests are written by hand. Automating test generation is difficult for several reasons. Writing correct expected wire-level assertions requires deep knowledge of the protocol. It is hard to distinguish a wrong test from a driver bug without human judgment. Spec authors need to understand why a test passes before they can trust it; an opaque generated test provides little confidence. These barriers have kept manual authoring the status quo. However, property-based testing or LLM-assisted generation---using the prose spec and existing tests as context---are promising directions we intend to explore.

\section{Related Work}

Our task---implementing the same network protocol and a consistent API in many programming languages---falls within the category of \textit{multilanguage development}~\cite{mussbacher_polyglot_2024}. The subcategory of \textit{polyglot programming} describes the development of an individual program with components in multiple languages; testing such programs is a well-studied problem~\cite{houdaille_polyglot_2024,yang_multi-language_2024}. However, our problem is to test the conformance of \textit{multiple} programs, each in a different language, to each other and to a specification. This subject has not been researched thoroughly, to our knowledge.

\subsection{Domain-Specific Test Languages}

Our YAML tests are an example of a \textit{domain-specific test language}, or DSTL~\cite{felicio_rapitest_2023,da_silva_test_2019,dragoi_domain_2023,ye_automated_2021,smithy_authors_http_nodate}. Many telecommunication protocols' test suites are written in the TTCN family of DSTLs~\cite{grabowski_design_2000}. A well-known DSTL is Gherkin, executed by Cucumber. We rejected Gherkin because it is designed for communication with non-technical stakeholders, whereas our tests are by and for engineers.

The Smithy project includes a DSTL that is more like our YAML tests~\cite{smithy_authors_http_nodate}. Smithy is an interface definition language for services. Engineers write test cases in Smithy's DSTL which specify exactly how client and server implementations should serialize remote procedure calls (RPCs) on the wire as HTTP messages. Similar to our YAML tests, Smithy's DSTL tests interact with both a client's ``upper'' and ``lower'' surface---the in-process API and the wire messages. Interpreters are available in at least nine languages.

Ecma International maintains Test262, containing over 50,000 handwritten test cases to check JavaScript implementations' conformance to the ECMA-262 standard~\cite{ye_automated_2021}. Like MongoDB, Ecma wrote a spec of the spec tests, describing the case file format and how test cases should be run. There are at least six independent implementations of the Test262 spec. COMFORT~\cite{ye_automated_2021} is an experimental system that uses an LLM to generate JavaScript test cases from ECMA-262. The LLM benefits from the spec's rigidly structured English. COMFORT's test cases do not include expected outputs; instead, COMFORT applies differential testing with many JavaScript implementations.

\subsection{Testing the Upper and Lower Surface}
\label{subsec:testing-the-upper-and-lower-surface}

A major strength of our approach is that we test not only the return values of API calls, but also the contents of drivers' messages on the wire. The ``ferry-clip'' test architecture is the closest antecedent of this idea~\cite{bochmann_protocol_1994,chanson_design_1989}. It tests an implementation of an OSI protocol layer by controlling the protocols above and below it in the OSI stack, attaching to the SUT's top and bottom like a ``clip.'' The test harness installs a proxy server which ``ferries'' messages produced by the SUT back to the test harness to be checked against assertions. Our tests' mechanics are quite different---we use direct procedure calls at the top surface, and command-monitoring callbacks at the bottom---but the architecture is analogous.

\subsection{Test Case Generation}
\label{test-case-generation}

MongoDB driver specifications are written in English, and we wrote our YAML tests by hand. But there are many examples of specifications that use formal logic or a finite state machine (FSM), from which one can generate test cases mechanically~\cite{richardson_approaches_1989,hoque_ensuring_2015,mcmillan_formal_2019,bishop_engineering_2006,singha_messi_2024,raghavan_protocol_1995,dssouli_test_1999}. We could model MongoDB's client-server system as an FSM and check that the FSM's transitions match the drivers'. We could also use an LLM to generate more test files from the prose spec and existing tests. The LLM could autonomously check its work: if every driver fails an LLM-made test, the test is probably wrong, but if only some fail, the test may have uncovered a bug.

\section{Conclusion}

We described a specification-based testing method developed over eleven years to ensure consistent behavior among libraries in a dozen languages, serving millions of users. We found that declarative, machine-interpretable test specs are effective at enforcing consistency. Our Unified Test Format minimizes the cost of maintaining tests: each driver implements a single UTF test runner; tests are written once and run everywhere. Our approach is not as advanced as some prior research, but its continuous use and evolution at scale offers an extended case study. We hope the lessons we learned prove useful to other teams enforcing consistent behavior across many implementations.

\bibliographystyle{IEEEtran}
\bibliography{references,other_references}

\end{document}